\documentclass[psfig]{aastex}

\usepackage{emulateapj5}

\def\ltsima{$\; \buildrel < \over \sim \;$}

\def\lsim{\lower.5ex\hbox{\ltsima}}
\def\gtsima{$\; \buildrel > \over \sim \;$}
\def\gsim{\lower.5ex\hbox{\gtsima}}

\begin{document}

\title{Self-Regulated Growth of Supermassive Black Holes in Galaxies
as the Origin of the Optical and X-ray Luminosity Functions of Quasars}

\author{J. Stuart B. Wyithe and Abraham Loeb}

\email{swyithe@isis.ph.unimelb.edu.au; loeb@sns.ias.edu}

\altaffiltext{1}{The University of Melbourne, Parkville, Vic, Australia}

\altaffiltext{2}{Institute for Advanced Study, Princeton, NJ 08540}

\altaffiltext{3}{Guggenheim Fellow; on leave from the Astronomy Department,
Harvard University, Cambridge, MA 02138}

\begin{abstract}
\noindent 

We postulate that supermassive black-holes grow in the centers of
galaxies until they unbind the galactic gas that feeds them.  We show
that the corresponding self-regulation condition yields a correlation
between black-hole mass ($M_{\rm bh}$) and galaxy velocity dispersion
($\sigma$) as inferred in the local universe, and recovers the
observed optical and X-ray luminosity functions of quasars at
redshifts up to $z\sim 6$ based on the hierarchical evolution of
galaxy halos in a $\Lambda$CDM cosmology. With only one free parameter
and a simple algorithm, our model yields the observed evolution in the
number density of optically bright or X-ray faint quasars between
$2\la z\la 6$ across 3 orders of magnitude in bolometric luminosity
and 3 orders of magnitude in comoving density per logarithm of luminosity.  
The self-regulation
condition identifies the dynamical time of galactic disks during the
epoch of peak quasar activity ($z\sim2.5$) as the origin of the
inferred characteristic quasar lifetime of $\sim10^7$ years.  Since
the lifetime becomes comparable to the Salpeter $e$-folding time at
this epoch, the model also implies that the $M_{\rm bh}-\sigma$
relation is a product of feedback regulated accretion during the peak
of quasar activity. The mass-density in black-holes accreted by that
time is consistent with the local black-hole mass density $\rho_{\rm
bh}\sim(2.3^{+4.0}_{-1.5})\times10^5M_\odot$Mpc$^{-3}$, which we have
computed by combining the $M_{\rm bh}$--$\sigma$ relation with the
{\em measured} velocity dispersion function of SDSS galaxies (Sheth et
al.~2003). Comparison of the local black-hole mass-function with that
inferred from combining the feedback-relation with the halo
mass-function suggests that most massive ($>10^9M_\odot$) black-holes
may have already been in place by $z\sim6$. Applying a similar self-regulation
principle to supernova-driven winds from starbursts, we find that the
ratio between the black hole mass and the stellar mass of galactic
spheroids increases with redshift as $(1+z)^{3/2}$ although the
$M_{\rm bh}$-$\sigma$ relation is redshift-independent.

\end{abstract}

\keywords{black-hole physics - quasars: general}

\section{Introduction}

It has long been recognized that the formation processes of supermassive
black-holes (BHs) and stars may be self-regulating (Dekel \& Silk 1986;
Silk \& Rees 1998), since the amount of energy released by quasars and
stars can exceed the binding energy of the gas clouds in which they
are born. Indeed, the stellar and central BH masses of
galaxies show striking correlations with the depth of the
gravitational potential well of their hosts.  These phenomenological
correlations take the form of the mass-velocity dispersion correlation
for BHs (Merritt \& Ferrarese 2001; Tremaine et al. 2002) and the
Tully-Fisher (1977) and Faber-Jackson (1976) relations for stars in
the present-day universe.

In this paper we explore the self-limiting growth of the BH and
stellar masses of galaxies as a function of cosmic time, and
investigate the consequences for the quasar luminosity function.
Recent work suggests that the kinetic output of quasars may be
comparable to their radiative output (see \S 2.1.1 in Furlanetto \&
Loeb 2001; Begelman~2003). Our simple approach postulates that growth
stops when the energy released by these sources exceeds the binding
energy of the gas within their host galaxies after one dynamical time
of the gas reservoir. Using the redshift-dependent abundance of
galactic halos as a function of potential depth in a $\Lambda$CDM
cosmology, we demonstrate that this simple approach describes the
luminosity function of quasars in optical and X-ray bands. We
adopt the WMAP set of cosmological parameters (Bennett et al. 2003),
namely mass density parameters of $\Omega_{m}=0.27$ in matter,
$\Omega_{b}=0.044$ in baryons, $\Omega_\Lambda=0.73$ in a cosmological
constant, a Hubble constant of $H_0=71~{\rm km\,s^{-1}\,Mpc^{-1}}$, an
rms amplitude of $\sigma_8=0.84$ for mass density fluctuations in a
sphere of radius $8h^{-1}$Mpc, and a primordial power-spectrum with a
power-law index $n=1$.

We begin the paper with a discussion of the regulation of BH growth by
feedback during luminous quasar phases (\S~\ref{feedback}) and then
discuss the implications for modeling of the optical
(\S~\ref{lumfunc}) and X-ray (\S~\ref{xraylf}) luminosity functions of
quasars, including the number counts of high-redshift X-ray quasars
(\S~\ref{numbercounts}). We subsequently compute the mass accreted
into BHs during luminous quasar phases (\S~\ref{accretion}) and
compare this to the local density in BHs (\S~\ref{density}), which we
compute using the measured velocity dispersion function of {\it Sloan
Digital Sky Survey} (SDSS) galaxies (Sheth et al.~2003). In
\S~\ref{density} we also demonstrate that the most massive BHs were
already in place at high redshift. We then show in \S~\ref{star} and
\S~\ref{magorrian} that while the $M_{\rm bh}-\sigma$ relation is
redshift-independent, the ratio of BH mass to stellar mass within
galaxies should increase with increasing redshift if both are
regulated by a similar feedback process. Finally we compare our
findings with some previous work in \S~\ref{prev} and summarize our
main conclusions in \S~\ref{conclusion}.

\section{Feedback and The Growth of Galactic Black-Holes}
\label{feedback}

The growth of a supermassive BH is expected to be accelerated during a
galaxy merger, when cold gas is driven to the center of the merger
remnant (Mihos \& Hernquist~ 1994; Hernquist \& Mihos~1995).  Yu \&
Tremaine~(2002) have found that the local mass density in BHs is
consistent with the integrated luminosity density of quasars if the
mean radiative efficiency of the accreting material is $\epsilon_{\rm
rad}\sim8\%$; moreover, they concluded that about half the current BH
mass density was accreted near the Eddington rate at redshift $z\ga 2$. It
is natural to expect that a quasar shining at its limiting Eddington
luminosity could generate a powerful galactic wind and eventually
terminate the accretion process that feeds it.

Indeed, Silk \& Rees~(1998) pointed out that winds from quasars could
generate self-regulating outflows in the surrounding gas, if the
energy in the outflow liberates as much as the binding energy of the
gas in a dynamical time.  For a quasar output power that scales as the
Eddington limiting luminosity, this condition leads to a power-law
scaling of the BH mass with halo velocity dispersion, $M_{\rm
bh}\propto \sigma^5$. This scaling was used by Haehnelt, Natarajan \&
Rees~(1998) to demonstrate consistency between the density of quasars
at $z\sim3$ and the time derivative of the Press-Schechter
mass-function for a quasar lifetime that is comparable with the
Salpeter time (the $e$-folding time of the BH mass), $\sim 2.4\times
10^7 (\epsilon_{\rm rad}/0.06)~{\rm yr}$. Here we extend an earlier
model (Wyithe \& Loeb~2002) to include feedback limited BH growth
within the context of hierarchical merging in a $\Lambda$CDM
cosmology. We compare our model with the most recent data for the
space density of high redshift quasars at both optical and X-ray
wavelengths.

We assume that following a merger, a BH shines at a fraction $\eta$ of
its Eddington luminosity ($\propto M_{\rm bh}$) returning fraction
$F_{\rm q}$ of this energy\footnote{If the energy output of the quasar 
is communicated to the galaxy through a wind with an outflow velocity that 
is not much larger than the escape velocity of the halo, then momentum 
conservation leads to the same condition as equation~(\ref{quasarbound}).} 
to the galactic gas.  The quasar will
unbind the galactic gas if it supplies as much as the gravitational
binding energy of the gas over the time it takes the gas to respond
dynamically to this energy injection, $t_{\rm dyn}$.  The cold gas is
usually located in a disk with a characteristic radius
$(\lambda/\sqrt{2})r_{\rm vir}\sim0.035r_{\rm vir}$ (Mo, Mao \&
White~1998), and hence a dynamical time\footnote{We assume that  
gas can be driven towards the black hole during the merger.
Such accretion is communicated over the characteristic dynamical time
$r/v$ (as appropriate for a radial orbit) rather than $2\pi r/v$ (as 
appropriate for a circular orbit).} of $t_{\rm dyn}\sim0.035r_{\rm
vir}/v_{\rm c}$.  Here $\lambda\sim 0.05$ is the spin parameter,
$r_{\rm vir}$ is the virial radius, and $v_{\rm c}$ is the circular
velocity of the galactic halo\footnote{One inaccuracy in our modeling
is that we assume dark matter halos to be singular isothermal spheres,
while an NFW profile gives a lower $v_{\rm c}$ at the disk radius than
at $r_{\rm vir}$.  We ignore this difference because galaxies appear
to have nearly flat rotation curves in practice.} (Barkana \& Loeb
2001),
\begin{equation}
v_{\rm c}=245~\left({M_{\rm halo}\over 10^{12}M_\odot}\right)^{1/3}
[\xi(z)]^{1/6}\left({1+z\over 3}\right)^{1/2}~
{\rm km~s^{-1}},
\end{equation}
where $\xi(z)$ is close to unity and defined as $\xi\equiv
[(\Omega_m/\Omega_m^z)(\Delta_c/18\pi^2)]$, $\Omega_m^z \equiv
[1+(\Omega_\Lambda/\Omega_m)(1+z)^{-3}]^{-1}$,
$\Delta_c=18\pi^2+82d-39d^2$, and $d=\Omega_m^z-1$ (see Barkana \&
Loeb 2001 for more details).
The value of $F_{\rm q}$ should be of order unity if the gas
traps much of the quasar output energy without radiating it away
(Begelman~2003).  The self-regulation condition is then,
\begin{equation}
\label{quasarbound}
\eta L_{\rm Edd,\odot} M_{\rm bh} F_{\rm q} = \frac{\epsilon_{\rm rad}
\delta M_{\rm bh}c^2}{t_{\rm dyn} } F_{\rm q}= 
\frac{\frac{1}{2}\frac{\Omega_{b}}{\Omega_{m}}M_{\rm
halo}v_{\rm c}^2}{t_{\rm dyn}},
\end{equation}
where $M_{\rm halo}$ is the halo mass, $\delta M_{\rm bh}$ is the mass
accreted and $L_{\rm Edd,\odot}$ is the Eddington luminosity per unit
mass ($M_\odot$). 
Note that we assume (as appropriate at high redshifts) that all the
gas within a galactic halo cools on a time much shorter than the
Hubble time.  Our primary objective is to show that this condition is
sufficient to describe the quasar luminosity function at $z\ga2$
(before groups and clusters of galaxies become prominent).
Equation~(\ref{quasarbound}) implies $M_{\rm bh}\propto v_{\rm c}^5$
as inferred for galactic halos in the local universe (Ferrarese
2002). The index of 5 is larger than the 4-4.5 inferred from the local
$M_{\rm bh}-\sigma$ relation (Merritt \& Ferrarese 2001; Tremaine et al. 2002), 
the difference having it origin in the observation that the $v_{\rm c}-\sigma$
relation is shallower than linear (Ferrarese~2002). The index of 5 is also 
larger than presented in the recent work of Baes, Buyle, Hau \& Dejonghe~(2003),
with the difference having its origin in the uncertainty in the slope of the
$M_{\rm bh}-\sigma$ relation.
Feedback regulated growth implies that the $M_{\rm bh}-v_{\rm
c}$ relation is independent of redshift,
\begin{equation}
M_{\rm bh}= 1.9\times10^8 M_\odot \left({\eta F_{\rm q}\over
0.07}\right)\left({v_{\rm c}\over 350~{\rm km~s^{-1}}}\right)^5.
\end{equation}
The redshift independence is consistent with the recent results of
Shields et al.~(2002). Using quasars out to $z\sim3$, they demonstrated 
that the $M_{\rm bh}-\sigma$ relation does not evolve with redshift.  
The resulting relation between BH and halo mass may be written as
\begin{eqnarray}
\label{eps}
\nonumber M_{\rm bh}(M_{\rm halo},z) &=& \epsilon M_{\rm halo}\\
&&\hspace{-25mm}= \epsilon_{\rm o}M_{\rm halo} \left(\frac{M_{\rm
halo}}{10^{12}M_{\odot}}\right)^{\frac{2}{3}}
[\xi(z)]^\frac{5}{6}(1+z)^\frac{5}{2}.
\end{eqnarray}
Assuming $\eta=1$ as suggested by observations of low and
high-redshift quasars (Floyd~2003; Willott, McLure \& Jarvis~2003), we
can compute the normalization in equation~(\ref{eps}) from
equation~(\ref{quasarbound}). The only adjustable parameter in our
model if $\eta\sim1$ is $F_{\rm q}$.  If the BH growth was mostly
complete by a redshift $z\sim1$-2 (Yu \& Tremaine~2002) so that the
local BH masses reflect the limiting values at that redshift, then a
value of $\epsilon_{\rm o}=10^{-5.7}$ 
is consistent with BH masses that are measured locally (Ferrarese
2002).  This implies a value of $F_{\rm q}=0.07$.

Modification of the simple scheme described above would be required
if cooling of the gas heated by the outflow is important so that
additional energy is required to unbind the gas. The factor required
could be as large as $c/v_{\rm c}$ based on momentum conservation
(Begelman~2003), with the resulting dependence $M_{\rm bh}\propto
v_{\rm c}^4$. This relation is shallower than observed.  The
importance of cooling can be evaluated by estimating the cooling time
for virialized gas. The cooling rate for primordial gas of temperature
$T_{\rm vir}\ga10^6$K and proton density $1.2\times
10^{-3}\left[({1+z})/{3}\right]^3$cm$^{-3}$ (typical for a virialized
halo) is $\Lambda\sim 8.7\times10^{-30}\left[({1+z})/{3}\right]^6
(T_{\rm vir}/10^6{\rm K})^{1/2}$erg cm$^{-3}$ s$^{-1}$ (Barkana \&
Loeb~2001). We have adopted a cooling rate that increases as
$\sqrt{T}$ above $T_{\rm vir}=10^6$K assuming that it is dominated by
bremsstrahlung at higher temperatures. The thermal energy per unit
volume in this gas is $E\sim
5\times10^{-13}\left[({1+z})/{3}\right]^3\left({T_{\rm
vir}}/{10^6\mbox{K}}\right)$erg cm$^{-3}$.  The cooling time for
virialized gas is therefore
\begin{equation}
t_{\rm cool}\sim\frac{E}{\Lambda}\sim
2\times10^9\left(\frac{1+z}{3}\right)^{-3}\left(\frac{T_{\rm
vir}}{10^6\mbox{K}}\right)^{1/2}\mbox{yr},
\end{equation}
where the virial temperature is given by (Barkana \& Loeb 2001),
\begin{equation}
T_{\rm vir}=2.2 \times 10^6\left({M_{\rm halo}\over
10^{12}M_\odot}\right)^{2/3} [\xi(z)]^{1/3} \left({1+z\over 3}\right)~
{\rm K}.
\end{equation}
The cooling time should be compared with the characteristic lifetime
of the quasar, which our model relates to the dynamical time of the
galactic disk,
\begin{equation}
t_{\rm dyn}=0.035 {r_{\rm vir}\over v_{\rm c}}=
10^7~[\xi(z)]^{-1/2}\left({1+z\over 3}\right)^{-3/2}~{\rm yr}.
\label{eq:life}
\end{equation}
Thus we expect the halo cooling time to be longer than $t_{\rm dyn}$
in the range of redshifts and virial temperatures relevant for bright
quasars. 
As soon as the gas is expelled into the galactic halo and is
heated beyond its virial temperature, it will not be able to cool
during the quasar activity. Hence we assume that accretion is halted
as soon as the quasar supplies the entire galactic gas with more than
its binding energy.

\section{Modeling the quasar luminosity function}

\begin{figure*}[htbp]
\epsscale{1.6} \plotone{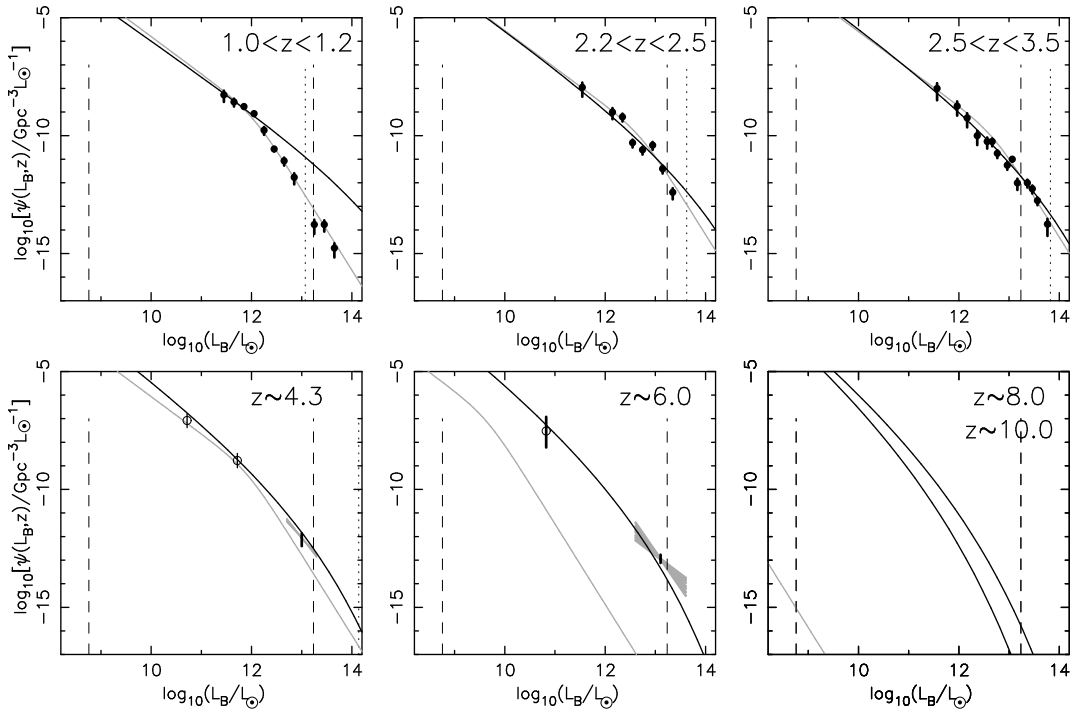}
\caption{\label{fig1} Comparison of the observed and model luminosity
functions. The data points at $z\la4$ are summarized in Pei~(1995), while
the light lines show the double power-law fit to the 2dF quasar luminosity 
function (Boyle et al.~2000).  At $z\sim4.3$ and
$z\sim6.0$ the data is from Fan et al.~(2001a;2001b;2003). The grey 
regions show the 1-$\sigma$ range of logarithmic slope
($[-2.25,-3.75]$ at $z\sim4.3$ and $[-1.6,-3.1]$ at $z\sim6$), and the vertical bars
 show the uncertainty in the normalization. The open circles show data
points converted from the X-ray luminosity function (Barger et al.~2003)
of low luminosity
quasars using the median quasar spectral energy distribution.
In each panel the vertical dashed lines correspond to the
Eddington luminosities of BHs bracketing the observed range of
the $M_{\rm bh}$-$v_{\rm c}$ relation, and the vertical dotted line
corresponds to a BH in a $10^{13.5}M_\odot$ galaxy.}
\end{figure*}
The peak in the quasar luminosity function at $z\sim2$ coincides with
the transition of the characteristic mass scale of collapsed halos
through the mass range of interest for the $M_{\rm bh}-M_{\rm halo}$
relation. In the present-day universe, the characteristic $1$-$\sigma$ 
mass scale is $\sim10^{13}$-$10^{14}M_\odot$. However galaxies reflect the
characteristic mass scale of $\sim10^{11}$-$10^{12}M_\odot$ at
$z\sim2$, and combine to make groups and clusters rather than
super-galaxies today. It may be that galaxies of this characteristic
size do not form due to prevention of cooling through conduction or
the removal of cold gas from halos (e.g. Benson et al.~2003), or
because the stars in giant ellipticals have already formed and
expelled the surrounding gas at $z\sim 6$ (Loeb \& Peebles~2003). The
excursion set formalism describes the rate of galaxy mergers near the
characteristic mass at times earlier than $z\sim2$, but describes the
formation rate of groups and clusters near the characteristic mass at
later times. The evolution of the bright end of the quasar luminosity
function is therefore only tractable to analytic descriptions at early
times ($z\ga2$), while the faint end should be tractable at all
redshifts. Since the observed quasar luminosity function implies that
around half the mass in BHs was accreted before $z\sim2$, the
$M_{\rm bh}-M_{\rm halo}$ relation should be evident in the bright end
of the quasar luminosity function at $z\ga2-3$. We now demonstrate that
this is indeed the case.

\subsection{The optical luminosity function}
\label{lumfunc}

We model the quasar luminosity function assuming that quasar activity
follows major galaxy mergers (mass ratios greater than 3) during which
BHs coalesce (Kauffmann \& Haehnelt~2000).  A linear relation between
BH mass and halo mass ($M_{\rm bh}\propto v_{\rm c}^3$) would be a
natural consequence of BH growth that is dominated by coalescence
(Haehnelt et al. 1998). However, a nonlinear relation must result from
significant mass accretion during the active quasar phase as found by
Yu \& Tremaine~(2002).

Our approach is based on the formalism described in Wyithe \&
Loeb~(2002).  Each quasar is assumed to shine at its Eddington
luminosity over a lifetime $t_{\rm q}$.  The quasar light curve may
therefore be written as
\begin{equation}
f(t)=\frac{L_{\rm Edd,B}}{M_{\rm bh}}\Theta\left(t_{\rm q}-t\right).
\end{equation} 
where $L_{\rm Edd,B}=5.73\times10^3M_{\rm bh}L_\odot$ is the Eddington
luminosity in the B-band of a BH with mass $M_{\rm bh}$ in solar
masses.  The quasar luminosity function (comoving number density of
quasars per B-band luminosity $L_{\rm B}$) is then
\begin{eqnarray}
\label{LF1}
\nonumber &&\hspace{-3mm}\Psi(L_{\rm B},z) = \\ \nonumber
&&\hspace{-3mm}\int_{0}^\infty dM_{\rm bh} \int_{0.25 \epsilon M_{\rm
halo}}^{0.5 \epsilon M_{\rm halo}} d\Delta M_{\rm bh} \int_z^\infty
dz' \left.\frac{dn_{\rm bh}}{dM}\right|_{\rm M=M_{\rm bh}-\Delta
M_{\rm bh}}\\ &&\hspace{-3mm}\times\left.\frac{d^2N_{\rm
merge}}{d\Delta M_{\rm bh}dt'}\right|_{\rm M_{\rm bh}-\Delta M_{\rm
bh}} \frac{dt'}{dz'} \delta\left[L_{\rm B}-M_{\rm bh}f(t_{\rm
z}-t')\right],
\end{eqnarray}
where $\frac{dn_{\rm bh}}{dM_{\rm bh}}$ 
and $\frac{d^2N_{\rm merge}}{d\Delta M_{\rm bh}dt'}$ are the mass
function and merger rate for BHs respectively, $t_{\rm z}$ is the
cosmic time at redshift $z$, and $\delta(x)$ is the Dirac delta
function.  Integrating over $M_{\rm bh}$ we find
\begin{eqnarray}
\nonumber &&\hspace{-3mm}\Psi(L_{\rm B},z) = \\ \nonumber
&&\hspace{-3mm}\int_{0.25 M_{\rm halo}}^{0.5 M_{\rm halo}} d\Delta
M_{\rm halo} \int_{\rm a_{\rm min}=\frac{1}{1+z_{\rm
form}}}^{a=\frac{1}{1+z}}da \left.\frac{dn_{\rm ps}}{dM}\right|_{\rm
M=M_{\rm halo}-\Delta M_{\rm halo}}\\
&&\hspace{-3mm}\times\left.\frac{d^2N_{\rm merge}}{d\Delta M_{\rm
halo}dt}\right|_{\rm M_{\rm halo}-\Delta M_{\rm halo}}
\frac{3}{5\epsilon}\frac{dt}{da}\frac{M_{\rm bh}}{L_{\rm Edd,B}},
\end{eqnarray}
where $a=1/(1+z)$ is the scale factor, $z_{\rm form}$ is the formation redshift 
of the quasar, and we have used the relation between 
black-hole and halo mass.
Evaluation of this expression requires the number of
mergers of halos of mass between $\Delta M$ and $\Delta M+d\Delta M$ 
with halos of mass $M$, $\frac{d^2N_{\rm merge}}{d\Delta M_{\rm halo} dt}|_{\rm M_{\rm halo}}$ 
(Lacey \& Cole~1993), as well as the co-moving density of halos with
mass $M$, $\frac{dn_{\rm ps}}{dM_{\rm halo}}$ (Press \& Schechter~1974).

Assuming $t_{\rm q}\ll H^{-1}(z)$, we then obtain $a-a_{\rm
min}=\frac{da}{dt}t_{\rm q}$, yielding the luminosity function\footnote{Our 
model does not account for the population of obscured quasars
(Fabian 2003). We implicitly assume that the powerful quasars detected at
high redshifts are capable of removing any surrounding gaseous torus that
may obscure them, since they release sufficient energy to unbind the gas
within the entire host galaxy.}
\begin{eqnarray}
\label{LF}
\nonumber &&\hspace{-3mm}\Psi(L,z)=\int_{0.25 M_{\rm halo}}^{0.5
M_{\rm halo}}d\Delta M_{\rm halo}\frac{3}{5\epsilon}\frac{t_{\rm
q}}{5.73\times10^3L_\odot M_\odot^{-1}}\\ &&\times\left.\frac{dn_{\rm
ps}}{dM}\right|_{\rm M=M_{\rm halo}-\Delta M_{\rm
halo}}\left.\frac{d^2N_{\rm merge}}{d\Delta M_{\rm
halo}dt}\right|_{\rm M_{\rm halo}-\Delta M_{\rm halo}},
\end{eqnarray}
where we have used $M_{\rm bh}={L_{\rm Edd,B}}/({5.73\times10^3L_\odot M_\odot^{-1}})$. 
We impose a lower cutoff on the mass of halos that can contain a
quasar.  Following reionization, the cutoff mass corresponds to the
Jeans mass in an ionized intergalactic medium (IGM), i.e. a virial
temperature of $\sim2.5\times10^{5}$K (Barkana \& Loeb 2001).
Equation~(\ref{quasarbound}) implies that the quasar lifetime should
be identified with the dynamical time of a galactic disk, $t_{\rm
q}\sim t_{\rm dyn}$, where $t_{\rm dyn}$ is given in
equation~(\ref{eq:life}).

We show a comparison between the observed and theoretical luminosity
functions in Fig.~\ref{fig1}.  The model successfully describes the
known features of the optical quasar luminosity function at
$z\ga2$. We also show on these plots the range of quasar luminosities
corresponding to BH masses observed locally to define the $M_{\rm
bh}-M_{\rm halo}$ relation (dashed lines). The black holes powering
quasars at $z\sim2$ are at the upper end of the mass range of the BHs
observed locally. As discussed above, the $M_{\rm bh}-M_{\rm halo}$
relation is obeyed locally by galaxies, but not by groups or clusters
of galaxies inside of which the hot gas is unable to feed the
BH. Our model luminosity function therefore over-predicts the number
of bright quasars at low redshift. However, our luminosity function
correctly describes the density of quasars at $z\sim1$ below the break
in the luminosity function. This luminosity regime corresponds to most
of the BH mass range in the locally observed $M_{\rm bh}-M_{\rm halo}$
relation. In order to match the bright end of the observed galaxy
luminosity function, the infall of gas into galaxies with $v_{\rm
c}\ga500~\mbox{km}\,\mbox{s}^{-1}$ must be suppressed (Kauffmann \&
Haehnelt~2000).  This scale indeed corresponds to the cut-off in the
$v_{\rm c}$ distribution of galaxies (Sheth et al. 2003). The
dotted lines show the luminosity of a BH within a galaxy of mass
$M_{\rm halo}\sim10^{13.5}$ corresponding to a circular velocity of
$\sim500\mbox{km}\,\mbox{s}^{-1}$ at $z=0$. This scale is close to
the break in the quasar luminosity function at low-redshifts.

\begin{figure*}[htbp]
\epsscale{1.65}
\plotone{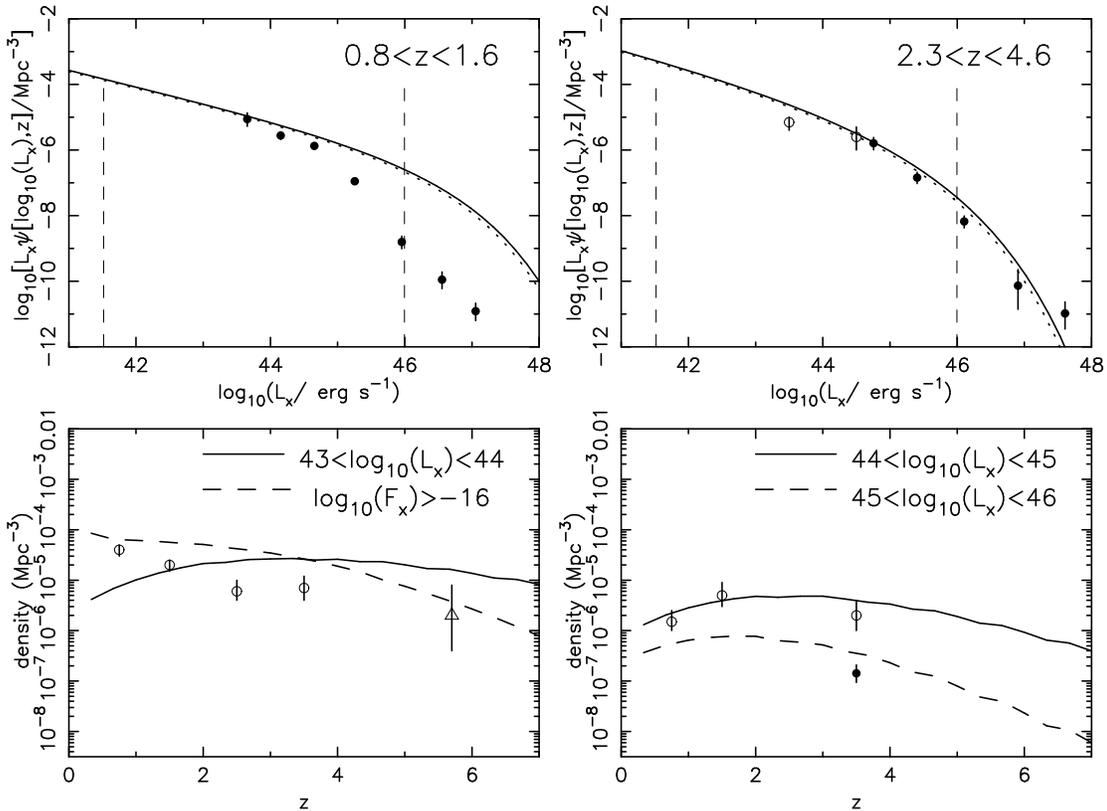}
\caption{\label{fig2} Upper panels: Comparison of the observed and
model X-ray luminosity functions. In each panel the vertical dashed
lines correspond to the Eddington luminosities of BHs bracketing the
observed range of the $M_{\rm bh}$-$v_{\rm c}$ relation. Lower
panels: evolution of the luminosity function with redshift in
different luminosity ranges. The filled data points are from the
luminosity function of Miyaji et al.~(2001), while the open circles
show results from Barger et al.~(2003). }
\end{figure*}
Near the peak of quasar activity at redshifts of $z\sim2$--$3$ our
model predicts a quasar lifetime of $t_{\rm dyn}\sim1-3\times10^7$
years.  This result naturally explains observational estimates of
quasar lifetimes (e.g Steidel et al.~2002; Yu \& Tremaine~2002; Haiman
\& Hui~2001; Martini \& Weinberg~2001). It is important to note that
our model suggests that the observed lifetime is typical of quasars
near the peak of the luminosity function, but that quasars at higher
redshift have shorter lifetimes.

At redshifts below the peak of quasar activity our model successfully
describes the luminosity function at luminosities fainter than the
break (the agreement of the slope is particularly good). Boyle et
al.~(2000) found that the luminosity function at $z\la2.5$ is well
described by a luminosity evolution. However inspection of
Fig.~\ref{fig1} suggests that the evolution of quasars does not follow
a luminosity evolution at high redshifts. In particular, our model
suggests that rather than being an intrinsic feature of the quasars
themselves, the break in the luminosity function is present only at
low redshift and is a product of the inability of gas to cool inside
massive dark matter halos.  We also see that the faint end slope
(reproduced at low redshift) evolves, becoming steeper at high
redshift.

We emphasize that the redshift dependence of $M_{\rm
bh}\propto(1+z)^{5/2}$ in equation~(\ref{eps}) is necessary to explain
the density of bright quasars at $z\sim6$. If the characteristic
feedback time was constant, so that $M_{\rm bh}\propto(1+z)$ as is the
case for stars (see \S 3), then the sharp decline in the
Press-Schechter (1974) mass function with increasing halo mass would
have resulted in a deficit of luminous quasars that cannot be
compensated for by arbitrarily long quasar lifetimes.  For example we
find that the predicted density of quasars at $z\sim2$-$3$ can be made
consistent with observations by using a quasar lifetime of $10^8$
years, but that the quasar abundance at $z\sim6$ is underestimated by
$\sim4$ orders of magnitude.  

The success of our model is particularly impressive, given the fact
that its only free parameter ($\eta F_{\rm q}$) 
is directly set by local observations of BH masses, and that it
assumes only that BH growth in a galactic disk is self regulating. In
the following section, we demonstrate that this success extends into
the X-ray regime.

\subsection{The X-ray luminosity function}
\label{xraylf}

Haiman \& Loeb~(1999) noted that the luminosity function of quasars in
the X-ray band at $z\sim3$ can be related to the B-band luminosity
function at that redshift if the median quasar spectral energy
distribution is universal (Elvis~1994). They then extrapolated their
model to predict the number counts of $z>5$ quasars and found that
$\sim50$ should be detected in a {\em Chandra} deep field. Barger et
al.~(2002) have since found one $z>5$ quasar in a circle of radius
$6'$ at a flux level $F_{\rm X}>2\times10^{-16}~{\rm erg~s^{-1}}$. The
new constraints provided by these very high redshift X-ray quasars (as
well as optical high redshift quasars) motivate us to revisit the
issue of modeling the X-ray luminosity function.

The number counts of X-ray quasars provide strong additional
constraints on models of the luminosity function. At high
luminosities, wide field surveys by the ROSAT satellite provide the
X-ray luminosity function in the same luminosity range as the optical
luminosity function.  Therefore at $z\sim3$, the comparison of the
model optical and X-ray luminosity functions with the respective data
sets tests the applicability of the median quasar spectrum adopted for
the modeling (Haiman \& Loeb~1999).  However the deep X-ray surveys
now becoming available (Cowie et al.~2002; Barger et al.~2003) probe
bolometric quasar luminosities at $z\sim3-5$ that are $\sim2$ orders
of magnitude fainter than are probed in the optical.  This large
dynamic range in luminosities provides a long lever arm for testing
models of quasar evolution.

\begin{figure*}[htbp]
\epsscale{.8}
\plotone{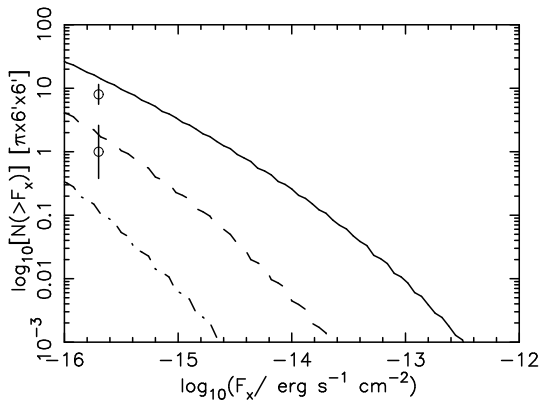}
\caption{\label{fig3} Number counts of quasars in a circle of radius
$6^\prime$ and redshifts above 3, 5 and 7 (top to bottom)
as a function of the flux limit. The number counts from Barger et
al.~(2003) at $z>3$ and $z>5$ with flux $>2\times10^{-16}~{\rm
erg~s^{-1}~cm^{-2}}$ are shown for comparison. }
\end{figure*}

We follow Haiman \& Loeb~(1999) and construct the X-ray luminosity
function in analogy with the B-band luminosity function using the
median quasar spectrum of Elvis~(1994). The universality of the spectrum 
is supported by the latest X-ray data on high redshift quasars 
(Brandt et al.~2002; Vignali et al.~2003).
In order to facilitate comparison with observed luminosity functions
we compute the luminosity $L_{\rm 2-8}$ in an (intrinsic) X-ray band
(e.g. 2--8keV) by integrating over the median spectrum. In analogy
with equation~(\ref{LF}) we find
\begin{eqnarray}
\label{LF}
\nonumber
&&\hspace{-8mm}\Psi(L_{\rm 2-8},z)=\int_{0.25 M_{\rm halo}}^{0.5 M_{\rm halo}}d\Delta M_{\rm
halo}\frac{3}{5\epsilon}\frac{t_{\rm q}}{L_{2-8,\odot}}\\
&&\times\left.\frac{dn_{\rm
ps}}{dM}\right|_{\rm M=M_{\rm halo}-\Delta M_{\rm
halo}}\left.\frac{d^2N_{\rm merge}}{d\Delta M_{\rm halo}dt}\right|_{\rm
M_{\rm halo}-\Delta M_{\rm halo}}.
\end{eqnarray}
Here $L_{2-8,\odot}=\int_{2\rm{keV}}^{8\rm{keV}} dE_\nu \nu L_{\nu,\odot}$,
where $\nu L_{\nu,\odot}$ is the spectral energy distribution of a solar mass
BH radiating at its Eddington limit and $E_\nu=h\nu$ is the photon
energy. 

In the upper two panels of Fig.~\ref{fig2} we show the predicted X-ray
luminosity function in the 0.5keV-2keV band (solid lines), and compare
it with the observed luminosity function derived by Miyaji et
al.~(2001) from ROSAT data (solid dots). Luminosity functions (solid
lines) are shown at $z\sim1$ and $z\sim3.5$. The bolometric
luminosities of the observed quasars are comparable with those of
bright optical quasars.  We also show results (open circles) for lower
luminosity X-ray AGN (Cowie et al.~2002; Barger et al.~2003). These
luminosities refer to the 2--8keV range and should be compared to the
dotted luminosity function. Several features of the comparison are
similar to those of the optical luminosity function.  The agreement is
excellent at redshifts above the peak in quasar activity. However
there is again an overproduction of bright quasars at $z\la 2$,
although the luminosity function below the break is reproduced. The
interpretation of these features was discussed in the previous
section.  As noted by Haiman \& Loeb~(1999) the success of the model
in producing both the optical and X-ray quasar luminosity functions at
$z\ga2$ indicates that the median quasar spectral energy distribution
is also a good representation at high redshifts. To help with the
orientation between the optical (Haiman \& Loeb~1998) and X-ray 
luminosity functions, we
have placed vertical dashed lines at luminosities corresponding to the
limits of the observed range of locally observed BH masses.

In the lower two panels of Fig.~\ref{fig2} we show the predicted
density of X-ray quasars (solid lines) of different luminosities as a
function of redshift in the 2keV-8keV band.  These are compared with
the observed densities derived by Cowie et al.~(2002) and Barger et
al.~(2003) at low luminosities (open circles) and Miyaji et al.~(2001)
at higher luminosities (solid dots). At $10^{43}$--$10^{44}~{\rm
erg~s^{-1}}$, the model under predicts the quasar density at
$z<1$. However at luminosities of $10^{44}$--$10^{45}~{\rm
erg~s^{-1}}$, the observed redshift evolution is well described by our
model.

We also compute the number counts of X-ray quasars. Barger et
al.~(2003) and Cowie et al.~(2002) have recently presented number
counts of faint, high redshift quasars in the {\em Chandra} deep field
north. The observations were made in the 0.4--6keV (observed) band and
the minimum flux level was $2\times10^{-16}~{\rm
erg~s^{-1}~cm^{-2}}$. We compute the comoving density of quasars with
observed fluxes above this level in this band as a function of
redshift by finding the luminosity function in the band
$0.4(1+z)$--$6(1+z)$keV. This density is shown as the dashed line in
the lower left panel of Fig.~\ref{fig2}, along with the density at
$z\sim5.7$ corresponding to the observation of one quasar in the
redshift range 5--6.5 within a circle of radius $6'$ (open
triangle). The model is in agreement with the observation of one source
at $z>5$.

\subsection{Number counts of high redshift X-ray quasars}
\label{numbercounts}

We can also compute the number counts of high redshift X-ray quasars
in the {\em Chandra} deep field north. Fig.~\ref{fig3} shows the
number of quasars expected from the model with redshifts above 3, 5
and 7, and within a circle of radius $6'$ as a function of the flux
limit. These are compared with the number counts from Barger et
al.~(2003) at $z>3$ and $z>5$. We find that $\sim 1$--$2$ quasars with
$z>5$ and flux $>2\times10^{-16}~{\rm erg~s^{-1}~cm^{-2}}$ are
expected in the {\em Chandra} deep field north compared with the one
observed. Similarly $\sim10$ quasars with $z>3$ and flux
$>2\times10^{-16}~{\rm erg~s^{-1}~cm^{-2}}$ are expected compared with
9 observed. This reproduction of the high redshift, low luminosity
number counts is an impressive success of our model. For example, the
Haiman \& Loeb (1999) model, which postulated a linear relation of
$M_{\rm bh} \propto M_{\rm halo}$, fails this test.

\begin{figure*}[htbp]
\epsscale{.8}
\plotone{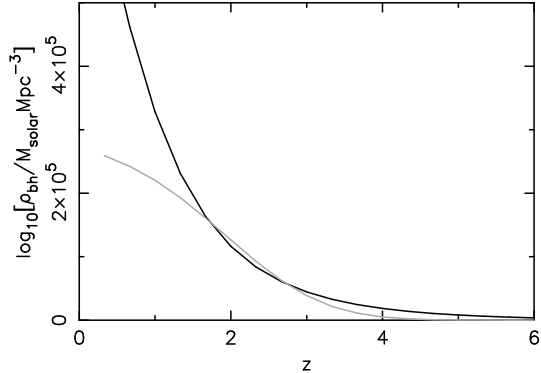}
\caption{\label{fig4} The mass density in supermassive BHs as a
function of redshift for the luminosity function computed in this
paper (dark line), and for the 2dF luminosity function (light line;
Boyle et al.~2000).}
\end{figure*}

Motivated by this success we can use the model to predict the number
counts of X-ray quasars at even higher redshifts. For example the
results of Fig.~\ref{fig3} suggest that $0.1$ quasar at $z>7$ would be
found in the {\em Chandra} deep field (or one quasar per 10
fields). Alternatively, probing deeper by a factor of $\sim3$ in flux
should allow detection of one quasar with $z>7$ per field.

\subsection{Using X-ray quasars to extend 
the optical luminosity function at high redshift to lower
luminosities}

The high redshift X-ray quasars discovered in the {\em Chandra} deep
field have bolometric luminosities that are 2 orders of magnitude
lower than the bright optically selected quasars.  Using the median
spectrum we can translate the density of X-ray quasars to points on
the optical luminosity function. In the $2.5<z<3.5$ and $z\sim4.3$
panels of Fig.~\ref{fig1} we plot the density of quasars with X-ray
luminosities between $10^{43}-10^{44}~{\rm erg~s^{-1}}$ and
$10^{43}-10^{44}~{\rm erg~s^{-1}}$, and $3\la z\la5$ 
in units of density per B-band
luminosity (open circles) for comparison with the model optical quasar
luminosity function. In the $z\sim6$ panel we plot the density
corresponding to one quasar per natural logarithm of luminosity in the
redshift range $5\la z\la6.5$.  The agreement is again excellent since
the comparison simply presents the same data in a different
way. However the comparison demonstrates agreement between model and
data over a large range of luminosities at high redshift. In
particular, it extends the comparison between model and data to the
lower logarithmic half of the observed range of BH masses. At
$z\sim4.3$ the model produces the normalization and slope of the SDSS
luminosity function, then becomes less steep at lower luminosities,
and intersects the point computed from the X-ray observations, some
2.5 orders of magnitude away in luminosity (and therefore BH mass) and
2.5 orders of magnitude away in density per logarithm of luminosity. 
Similarly, the model satisfies
the normalization of the luminosity function at $z\sim6$ for both the
low and high luminosity quasars as well as the constraint on the slope
of the bright end of the luminosity function. The model also
reproduces the suggestion from the data that the density of quasars
changes more rapidly at higher luminosities between $z\sim4.3$ and
$z\sim6$.

\section{accretion of black-hole mass during luminous phases}
\label{accretion}

The comoving mass density accumulated by quasar BHs by a redshift z is
\begin{equation}
\rho_{\rm bh}(z)=\int_z^\infty dz'\int_0^\infty dL_{\rm B} \frac{L_{\rm bol}}{\epsilon_{\rm rad}c^2}\Psi(L_{\rm B},z')\frac{dt}{dz},
\end{equation}
where $L_{\rm bol}$ is the bolometric luminosity of a quasar with
B-band luminosity $L_{\rm B}$, and $\epsilon_{\rm rad}$ is the
efficiency of conversion of accreted mass to radiation.  We take
$\epsilon_{\rm rad}=0.06$, appropriate for the last stable orbit of a
thin disk accretion onto a Schwarzschild BH. In Fig.~\ref{fig4} we
show the mass density as a function of redshift for the luminosity
function computed in this paper (dark line), and for the 2dF
luminosity function (light line; Boyle et al.~2000). We find that
$\rho_{\rm bh}\sim2\times10^5M_\odot\mbox{Mpc}^{-3}$ has been
accreted by $z\sim1.5$, similar to estimations of the local BH density
(e.g. Yu \& Tremaine~2002; Aller \& Richstone~2002).  Our luminosity
function model erroneously indicates that growth continues at $z<1.5$
due to the overestimate of massive galaxies at low redshift.  As
discussed in the analysis of Yu \& Tremaine~(2002), we find that an
appreciable fraction of the present mass density in supermassive BHs
is not accreted until after $z\sim3$. Yu \& Tremaine~(2002) also find
that $\sim90\%$ of the present-day BH mass density is accreted before
$z\sim1.5$. Indeed, our luminosity function suggests that the accreted
mass-density at $z\sim1.5$ is comparable to the local value, and
in agreement with the estimate from the 2dF luminosity function.  We
reiterate the point that by $z\sim2$ $t_{\rm dyn}$ is becoming
comparable to the Salpeter time. Within the scheme outlined in this
paper, that is the reason why most quasar growth occurs during
this period. We therefore suggest that it is the feedback regulated
accretion around the time of the peak in quasar activity that shapes
the $M_{\rm bh}-M_{\rm halo}$ relation.

\section{The Local Mass Function of Massive Black-Holes}
\label{density}

The local density of supermassive BHs can be computed by combining the
$M_{\rm bh}-\sigma$ relation with the velocity function of galaxies
(e.g. Yu \& Tremaine~2002; Aller \& Richstone~2002). Until recently
the velocity function had to be computed through combination of a
galaxy luminosity function with the Faber-Jackson~(1976) and
Tully-Fisher~(1977) relations. A more reliable representation is now
possible using the measured velocity dispersion function of early type
galaxies. Sheth et al.~(2003) presented the measured velocity
dispersion function for early type galaxies (which contain the most
massive BHs), as well as a model for the velocity dispersion function
of late-type galaxies.  They suggested an analytic fit of the form
\begin{equation}
\phi(\sigma)d\sigma =
\phi_\star\left(\frac{\sigma}
{\sigma_\star}\right)^\alpha
\frac{\exp{[-(\sigma/\sigma_\star)^\beta]}}
{\Gamma(\alpha/\beta)}\beta\frac{d\sigma}{\sigma},
\end{equation}
where $\phi_\star$ is the number density of galaxies and
$\sigma_\star$ is a characteristic velocity dispersion. Sheth et
al.~(2003) found that the parameters $\sigma_\star$, $\alpha$ and
$\beta$ are strongly correlated with one another. We take
$\phi_\star=(2.0\pm0.1)\times10^{-3}$Mpc$^{-3}$, $\alpha=6.5\pm1.0$,
$\beta=(14.75/\alpha)^{0.8}$ and
$\sigma_\star=161\Gamma(\alpha/\beta)/\Gamma[(\alpha+1)/\beta]~{\rm
km~s^{-1}}$.  From Fig.~6 in Sheth et al.~(2003) we model the velocity
dispersion function in the regime below $\sigma\sim200~{\rm
km~s^{-1}}$ (where it is dominated by late-type galaxies and
measurements are not provided by SDSS) using the functional form
\begin{equation}
\phi(\sigma)d\sigma =
\log_{10}(e)\times10^{1.73-1.71\log_{10}(\sigma)}\frac{d\sigma}{\sigma}.
\end{equation}
To compute the mass function of BHs, we combine $\phi(\sigma)$ with
the $M_{\rm bh}-\sigma$ relation
\begin{equation}
M_{\rm bh}=M_{\rm bh,200}\left(\frac{\sigma}{200~{\rm
km~s^{-1}}}\right)^\chi.
\end{equation}
Merritt \& Ferrarese~(2001) quote $M_{\rm
bh,200}=(1.48\pm0.24)\times10^8M_\odot$ and $\chi=4.65\pm0.48$.  We
computed a large number of realizations assuming that the parameters
$\phi_\star$, $\alpha$, $M_{\rm bh,200}$ and $\chi$ are distributed as
Gaussian within their quoted uncertainties.  We then find the mean and
variance of the logarithm of the resulting set of mass functions.

\begin{figure*}[htbp]
\epsscale{.8}
\plotone{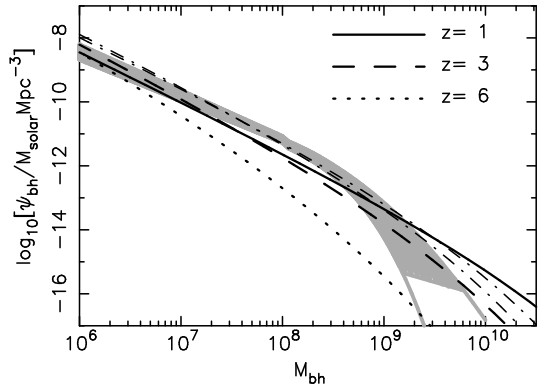}
\caption{\label{fig5} The mass function of supermassive BHs. The
grey region shows the mass function estimated from the velocity
function of Sheth et al.~(2003) and the $M_{\rm bh}-\sigma$ relation
of Merritt \& Ferrarese~(2001). The lower bound corresponds to the
lower limit in density for the observed velocity function 
($\sigma\frac{d\Phi}{d\sigma}>10^{-6.2}$), while the grey lines show
the extrapolation to lower densities. We also show the mass function
computed at $z=1$, 3 and 6 from the Press-Schechter~(1974) halo mass
function and equation~(\ref{eps}), as well as the mass function at
$z\sim2.35$ and $z\sim3$ implied by the observed density of quasars
and a quasar lifetime $t_{\rm dyn}$ (dot-dashed lines).}
\end{figure*}

In Fig.~\ref{fig5} we show the range ($\pm$ twice the variance) in the
computed mass function of BHs (solid area). This area is truncated below 
the region corresponding to the measured minimum density and the maximum 
measured velocity bin. The grey lines show the extrapolation to lower densities.
The derived BH mass
function differs substantially from that described in Aller \&
Richstone~(2002). In particular, the cutoff at the high mass end is
not as sharp in the present case. The reason is that the observed
velocity function does not agree with that derived from the galaxy
luminosity function and the Faber-Jackson relation (for a discussion,
see Sheth et al.~2003).

We have computed the local mass-density in BHs by integrating the
BH mass-function.  We find a local BH density of $\rho_{\rm
bh}=(2.3^{+4.0}_{-1.5})\times10^5M_\odot$Mpc$^{-3}$, where the
quoted error is the density computed from mass-functions at the
boundaries of the range shown in Fig.~\ref{fig5}. Using the
alternative values of $M_{\rm bh,200}=1.14\times10^{8.13\pm0.06}$ and
$\chi=4.02\pm0.32$ (Tremaine et al.~(2002) we find $\rho_{\rm
bh}\sim(2.5^{+4.0}_{-1.7})\times10^5M_\odot$Mpc$^{-3}$ These estimates
of BH density agree with the estimate of mass accreted during luminous
quasar phases as described by the 2dF luminosity function
(\S~\ref{accretion}).

The local BH mass function may be compared with the redshift dependent
mass function derived by combining the Press-Schechter~(1974) mass
function with the $M_{\rm bh}-M_{\rm halo}$ relation (\ref{eps}). We
have plotted the resulting mass function at $z=1$, 3 and 6 in
Fig.~\ref{fig5}. We find that the local density of BHs with masses
below $M_{\rm bh}\sim10^8M_\odot$ were in place by $z\sim1-3$,
coinciding with the peak in the quasar activity. More massive BHs were
in place at even higher redshifts.  In particular, we find that the
most massive BHs known ($M_{\rm bh}\sim3\times10^9M_\odot$) may have been
in place by $z\sim6$. This finding is in agreement with the conclusion of
Loeb \& Peebles~(2003) that stars in giant ellipticals have already
formed and expelled the surrounding gas at $z\sim 6$, but here the
conclusion is derived from a completely independent argument.

\subsection{The nearest very massive black-holes}

Consider a BH of mass $M_{\rm bh}\sim3\times10^9M_\odot$. We can
estimate the co-moving density of black-holes directly from the
observed quasar luminosity function and our estimate of quasar
lifetime.  At the peak of quasar activity, quasars powered by $M_{\rm
bh}\sim3\times10^9M_\odot$ BHs had a comoving density of $\Psi(>L)\sim
50\mbox{Gpc}^{-3}$.  However $t_{\rm hubble}/t_{\rm dyn}\sim 2\times
10^2$, so that the comoving density of BHs was
$\sim10^4\mbox{Gpc}^{-3}$. As this corresponds to the peak of BH
growth, the BH density should match the value observed today. The
density implies that the nearest BH of $\sim3\times10^9M_\odot$ should
be at a distance $d_{\rm bh}\sim
\left(4\pi/3\times10^4\right)^{1/3}\mbox{Gpc} \sim30\mbox{Mpc}$ which
is comparable to the distance of M87, a galaxy known to possess a BH
of this mass (Ford et al. 1994).

{\it What is the most massive BH that can be detected dynamically in a
local galaxy redshift survey?} The {\it Sloan Digital Sky Survey}
probes a volume of $\sim1\mbox{Gpc}^3$ out to a distance $\sim30$
times that of M87.  At the peak of quasar activity, the density of the
brightest quasars implies that there should be $\sim100$ BHs with
masses of $3\times10^{10}M_\odot$ per $\mbox{Gpc}^{3}$, the nearest of
which will be at a distance $d_{\rm bh}\sim130\mbox{Mpc}$, or $\sim 5$
times the distance to M87.  The radius of gravitational influence of
the BH scales as $M_{\rm bh}/v_{\rm c}^2\propto M_{\rm bh}^{3/5}$. We
find that for the nearest $3\times10^9M_\odot$ and
$3\times10^{10}M_\odot$ BHs, the angular radius of influence should be
similar.  Thus the dynamical signature of $\sim 3\times
10^{10}M_\odot$ BHs on their stellar host should be detectable.

The luminosities of the brightest quasars at $z\sim2$--$3$ imply
BH masses of $\sim10^{10}M_\odot$, even if they accrete at
their Eddington rate. The masses could be lower if the emission is
beamed, though this does not appear to be the case for high redshift
quasars (e.g. Willott, McLure \& Jarvis~2003).  The implied density
of $\sim10^{10}M_\odot$ BHs is in excess of the densities
obtained by convolving the galaxy velocity function (Sheth et al.~2003)
with the $M_{\rm bh}-\sigma$ relation (Merritt \& Ferrarese~2001; 
Tremaine et al.~2002). This is shown by the dot-dashed lines in
Fig.~\ref{fig5}. Here we have estimated the density of black-holes as
implied by the quasar luminosity function at $z\sim2.3$ and $z\sim3$
according to
\begin{equation}
\frac{d\Psi_{\rm bh}}{dM_{\rm bh}} = \frac{t_{\rm q}}{H^{-1}}\frac{dL}{dM_{\rm bh}}\Psi(L),
\end{equation}
where we have used $dM_{\rm bh}={dL}/{5.73\times10^3}$.
The presence of these BHs at $z\sim2-3$ may imply that these
very massive BHs will be found in the central galaxies of
large clusters of galaxies.

\section{Regulation of Star Formation}
\label{star}

We may derive the limiting stellar mass of a galaxy in view of the
same self-regulation principle (Dekel \& Woo~2002).  Suppose that
stars form with an efficiency $f_\star$ out of the gas that collapses and
cools within a dark-matter halo, 
and that a fraction $F_{\rm SN}$ of each supernova energy output,
$E_{\rm SN}$, heats the galactic gas mechanically. The mechanical
feedback will halt the star formation once the cumulative energy
returned to the gas by supernovae equals its binding energy (assuming
negligible radiative losses for a sufficiently rapid starburst). 
Hence the limiting stellar mass is set by the condition,
\begin{equation}
\label{starbound}
 \frac{M_{\star}}{w_{\rm SN}}
E_{\rm SN}F_{\rm SN} = E_{\rm
b}=\frac{1}{2}\frac{\Omega_{b}}{\Omega_{m}}M_{\rm halo}v_{\rm c}^2,
\end{equation}
where $w_{\rm SN}$ is the mass in stars (in $M_\odot$) per supernova
explosion.  Equation~(\ref{starbound}) implies that the total mass in
stars, $M_\star=(f_\star \Omega_b/\Omega_m)M_{\rm halo}$, scales as
\begin{equation}
\label{Mstar}
M_\star\propto M_{\rm
halo}^{5/3}\left[\xi(z)\right]^\frac{1}{3}\left(1+z\right).
\end{equation}

At a fixed redshift, the star formation efficiency scales as
$f_\star\propto v_{\rm c}^2 \propto M_{\rm halo}^{2/3}$. Thus smaller
galaxies are less efficient at forming stars, but a galaxy of fixed
mass is more efficient at higher redshift.  A Scalo~(1986) mass
function of stars has $w_{\rm SN}=126M_\odot$ per supernova and
$E_{\rm SN}=10^{51}$ ergs, and so we find that a mass in stars of
$M_\star=3\times10^{10}M_\odot$ and a velocity of $v_{\rm
c}\sim175\mbox{km}\,\mbox{s}^{-1}$ [the typical value observed locally
(Bell \& De Jong~2001)] implies $F_{\rm SN}\sim 0.5$ and
$f_\star=0.07$. The star formation efficiency in the most massive
spiral galaxies is therefore expected to be $\sim10\%$.

\section{Redshift Dependence of the Magorrian et al. (1998) Relation}
\label{magorrian}

In the preceding sections we have found that the mass of BHs
scales as $v_{\rm c}^5$ while the star formation efficiency scales
as $v_{\rm c}^2$. This is because the luminosity of a quasar is
limited by the Eddington value and is proportional to the BH
mass, while the energy release rate of supernovae is proportional to
the total stellar mass.  Hence we find that the ratio of BH to
stellar mass,
\begin{equation}
\frac{M_{\rm
bh}}{M_\star}\propto\left[\xi(z)\right]^\frac{1}{2}
\left(1+z\right)^\frac{3}{2},
\label{eq:Magorrian}
\end{equation} 
is independent of mass as observed (Magorrian et al. 1998). However,
we find that the relation evolves with redshift, and that the BH comprises a
larger fraction of the stellar mass at a higher redshift. This effect
has been seen in a sample of gravitationally lensed quasars (with
lensed host galaxies); Rix et al.~(1999) found that host galaxies are
much fainter at $z\sim2$ when compared with lower redshift galaxies
containing quasars of comparable luminosity. Our results are also
consistent with the semi-analytic predictions of Kauffmann \&
Haehnelt~(2000).

\section{Comparison with Previous Work}
\label{prev}

Since the work by Haehnelt, Natarajan \& Rees~(1998) there have been
several attempts to describe the quasar luminosity function within
hierarchical merging scenarios through semi-analytic
modeling. Kauffmann \& Haehnelt~(2000) assumed that the $M_{\rm
bh}$--$\sigma$ relation was set by the amount of gas that cools and
becomes available for star formation. They found that they could
explain the decrease in quasar activity at low redshift based on the
decrease in merger rates, the decrease in the gas available to fuel
BHs, and the assumption that gas accrets more slowly at late
times. For an accretion timescale that scales as $(1+z)^{-3/2}$, these
considerations led to quasar hosts that are a factor of 10 less
massive at $z\sim2$ than in the local universe.  Volonteri, Haardt \&
Madau~(2002) followed the evolution of the BH population through a
merger-tree algorithm. They modeled the luminosity function by fixing
the $M_{\rm bh}-\sigma$ relation from observations, and forcing the
quasar lifetime following a merger to be equal to the time for
accretion of the mass to match the $M_{\rm bh}-\sigma$ relation at an
efficiency of 10\%.

Di Matteo, Croft, Springel \& Hernquist~(2003) noted that the
above studies either assume the $M_{\rm bh}-\sigma$ relation from
observations, or derive it based on additional assumptions. They
suggested that a linear relationship between the gas in bulges and
$M_{\rm bh}$ results in the local $M_{\rm bh}-\sigma$ relation, the
drop in quasar density at low redshift, and the regulation of BH
growth through supernovae feedback. However their model over-predicts
the density of high-redshift quasars. In their conclusion, Di Matteo
et al.~(2003) state that it seems unlikely that BH accretion affects
gas dynamics in galaxies unless feedback from the quasar itself is
important.  On the other hand, the importance of quasar feedback has
recently been highlighted by Begelman~(2003).

In the previous sections, we have demonstrated that a scenario in
which quasar feedback limits BH growth through regulation of the
quasar lifetime naturally reproduces the quasar luminosity function at
high redshifts. For the first time we have compared the model to both
the $z\sim 6$ SDSS luminosity function and the X-ray luminosity
function, allowing comparison over $\sim3$ orders of magnitude in flux
and 3 orders of magnitude in quasar density per logarithmic luminosity 
interval at $z>5$.  The success of the model signals several important
implications. First, around the peak of quasar activity at $z\sim2.5$
the quasar lifetime, interpreted here as the dynamical time of the
host galactic disk, is $\sim10^7$ years [see
Eq. (\ref{eq:life})]. This provides a natural explanation for
observational estimates of the lifetime (e.g. Steidel et al.~2002; Yu \&
Tremaine~2002). Moreover, the dynamical time of the disk becomes
comparable to the Salpeter growth time at this redshift, so that
most of the mass in a given BH was added during this epoch. Thus we
suggest that the growth of BHs during luminous quasar phases around
this time, regulated by feedback from the quasar itself, set the slope
and amplitude of the the $M_{\rm bh}-v_{\rm c}$ relation observed
today.

\section{Conclusions}
\label{conclusion}

We have shown that if BHs grow in the centers of galaxies until they
unbind the galactic gas that feeds them over a dynamical time, then
the locally observed relation between BH mass and halo velocity
dispersion (Ferrarese 2002) is recovered both in slope and
normalization. We include this feedback regulation of BH growth in a
model of the quasar luminosity function and find that the observed
optical quasar luminosity function at $z\ga2$ as well as at
luminosities below the break at lower redshifts, are reproduced for
hierarchical merging of galactic halos in the standard $\Lambda$CDM
cosmology.  Moreover, using the median quasar spectral energy
distribution we showed that the same model yields good agreement with
the X-ray quasar luminosity function at $z\ga2$ (and the luminosity
function below the break at lower redshifts).  The model suggests that
$\sim10$ faint quasars at $z>3$ and $\sim1$ at $z>5$ should have been
found in the {\em Chandra} deep field north at a flux above
$2\times10^{-16}~{\rm erg~s^{-1}~cm^{-2}}$, in agreement with the
observed number counts. Overall, the model describes observations of
the high redshift luminosity function over $\sim3$ orders of magnitude
in bolometric luminosity and $\sim3$ orders of magnitude in comoving
density per logarithm of luminosity.  
This success is impressive given the fact that the model is
based on a simple algorithm with only one adjustable parameter, namely
the quasar feedback power in Eddington units $\eta F_{\rm q}$, and
that this quantity is set by local observations of BH masses.  Our
model associates the inferred quasar lifetime of $\sim 10^7$ years
(Steidel et al.  2002) with the dynamical time of a galactic disk
during the peak of quasar population ($z\sim2$--$3$).  Moreover, since
the quasar lifetime is comparable to the Salpeter time during this
epoch, the model suggests that the slope and normalization of the
locally observed $M_{\rm bh}-M_{\rm halo}$ relation resulted from
feedback regulated accretion around this epoch.  We have computed the
local mass-function of BHs by combining the $M_{\rm bh}-\sigma$
relation with the measured velocity dispersion function of Sheth et
al.~(2003). 
Comparison of the local BH mass function with the mass function
inferred from the feedback relation and halo mass-function implies
that the most massive galaxies already hosted their present-day
BHs by $z\sim3$--$6$ (see also Loeb \& Peebles 2003).  Finally, we
showed that if supernova feedback regulates star formation in a
similar way, then the Magorrian et al.  (1998) relation between the BH
and stellar masses of galactic spheroids should be redshift dependent
[see Eq.~(\ref{eq:Magorrian})], as observed for the hosts of
gravitationally lensed quasars (Rix et al.~1999).

\acknowledgements 

JSBW thanks the Institute for Advanced Study for its pleasant
hospitality during the course of this work.  AL acknowledges support
from the Institute for Advanced Study and the John Simon Guggenheim
Memorial Fellowship.  This work was also supported in part by NSF
grants AST-0071019, AST-0204514 and NASA grant ATP02-0004-0093.

\end{document}